\documentclass[12pt]{article}
\usepackage[pdftex]{graphicx}
\usepackage{amssymb}
\usepackage{amsmath}
\usepackage{bm}
\usepackage{enumerate}
\usepackage{xcolor}

\newcommand{\Tr}{\mbox{Tr}}

\begin{document}

\begin{titlepage}
	\vskip 2cm
	\begin{center}
		\Large{{\bf On Strongly Coupled Matrix Theory\\and Stochastic Quantization:\\A New Approach to Holographic Dualities}}
	\end{center}

	\vskip 2cm
	\begin{center}
		{Nick Heller\footnote{\tt{nheller@hmc.edu}}, Vatche Sahakian\footnote{\tt{sahakian@hmc.edu}}}
	\end{center}
	\vskip 12pt
	\centerline{\sl $\mbox{}^{1,2}$ Harvey Mudd College}
	\centerline{\sl Physics Department, 241 Platt Blvd.}
	\centerline{\sl Claremont CA 91711 USA}
	\vskip 12pt
	\centerline{\sl $\mbox{}^{2}$ Institut de Physique Th\'{e}orique}
	\centerline{\sl Universit\'{e} Paris Saclay, CEA, CNRS}
	\centerline{\sl Orme des Merisiers, 91191 Gif-sur-Yvette Cedex, France}

	\vskip 1cm
	\begin{abstract}
	Stochastic quantization provides an alternate approach to the computation of quantum observables, by stochastically sampling phase space in a path integral. Furthermore, the stochastic variational method can provide analytical control over the strong coupling regime of a quantum field theory -- provided one has a decent qualitative guess at the form of certain observables at strong coupling. In the context of the holographic duality, the strong coupling regime of a Yang-Mills theory can capture gravitational dynamics. This can provide enough insight to guide a stochastic variational ansatz. We demonstrate this in the bosonic Banks-Fischler-Shenker-Susskind Matrix theory. We compute a two-point function at all values of coupling using the variational method showing agreement with lattice numerical computations and capturing the confinement-deconfinement phase transition at strong   coupling. This opens up a new realm of possibilities for exploring the holographic duality and emergent geometry.
	\end{abstract}
\end{titlepage}

\newpage \setcounter{page}{1}

\section{Introduction and highlights}
\label{sec:intro}

In the context of holographic dualities~\cite{Banks:1996vh}-\cite{Itzhaki:1998dd}, one can use a non-gravitational theory -- typically a Yang-Mills theory of rank $N$ and effective coupling $\lambda$, to describe gravitational dynamics in a regime where $N$ is large and $\lambda>1$. While large $N$ can simplify the treatment, the strong coupling problem of gravitational holography is a significant obstacle to our understanding of both gravity and the holographic map. In practice, usually the best one can do is to utilize symmetry to study constrained or protected observables, or to use numerical lattice techniques to access the strong coupling regime of the non-gravitational theory.

In this work, we take a first step in tackling this strong coupling problem with a different approach. For concreteness, we focus on the Banks-Fischler-Shenker-Susskind (BFSS) Matrix theory~\cite{Banks:1996vh}, which describes M-theory through a Matrix model that is the dimensional reduction of $10$D $\mathcal{N}=1$ Super Yang-Mills (SYM). In particular, at low energies, eleven-dimensional supergravity in the light-cone frame can be described by $0+1$ dimensional $SU(N)$ matrix quantum mechanics in the regime where the effective coupling and $N$ are large\footnote{The correspondence can also be extended to finite $N$ and Discrete Light-Cone Quantized (DLCQ) M-theory~\cite{Seiberg:1997ad,Bigatti:1997gm}.}. This quantum mechanics arises from the dynamics of D0 branes and involves a super-renormalizable interaction with coupling $g$ that carries units of energy cubed. At large $N$, the effective 't Hooft coupling at energy scale $E$ is $\lambda=g^2 N/E^3$. The bosonic version of this theory, studied numerically at finite temperature, is known to exhibit a confinement/de-confinement phase transition that occurs at $\lambda\sim 1$~\cite{Filev:2015hia,Tanwar:2020fuv}\footnote{See also  key earlier work on numerical BFSS in~\cite{Catterall:2007fp,Anagnostopoulos:2007fw}}. This phase transition is absent in the supersymmetric version of the theory. Our goal is to reproduce the numerical results from the bosonic BFSS theory using new analytical techniques that allow us to directly access strongly coupled dynamics. 

Stochastic quantization~\cite{Parisi:1980ys,Damgaard:1988nq} is a well-established approach that has been studied for decades. It allows one to think about the phase space explored by a system inside a quantum path integral as a stochastic evolution of the Langevin type. The fields are made to evolve stochastically in such a way that equilibrium is achieved at the saddle point of the action; as a result, statistical fluctuations at equilibrium explore the relevant phase space that dominates the quantum path integral. It has been shown in various settings, from scalar field theories to non-abelian gauge theories~\cite{Namiki:1982mj}, that this approach can be used to compute quantum vacuum expectation values (vevs) of operators perturbatively using statistical averaging in a stochastic ensemble. Much more interestingly, stochastic quantization offers in addition a variational approach to computing such observables~\cite{Grandati:1992hj}-\cite{Greensite:1982ck}, one that can be applied analytically in strongly coupled regimes. The key to the variational approach is to have a good ansatz for an observable. In traditional quantum field theories, this can be a very difficult task. However, in holographic dualities, due to knowledge of the dual gravitational side, good guesses can be constructed. For example, BFSS theory is known to be chaotic and, at strong coupling, can be described by random matrices with a coupling-dependent effective mass. We will use this knowledge to employ a robust ansatz and access strong coupling dynamics.

We consider the BFSS Matrix model without the fermions using only the $N\times N$ matrix bosonic fields $X^1(\tau),\ldots,X^D(\tau)$ in the adjoint of $SU(N)$ -- where $D\leq 9$, and considering also the addition of a mass term for these fields to lift flat directions that arise from commuting matrix configurations. We measure the effective coupling of the theory at the scale of this mass. In this work, we develop the stochastic quantization of this Matrix theory in detail. First, we employ a perturbative approach in the effective coupling, deriving stochastic Feynman diagrams to compute any observable. We use this perturbative approach to compute the vev of $R^2\equiv \mbox{Tr}\,{X^i}^2/N$, which is a gauge invariant operator measuring the size of the matrix configuration in the vacuum. This operator is known from earlier numerical computations to signal phase transition that cannot be seen perturbatively. Next, we proceed to developing the variational approach of the stochastic quantization of bosonic BFSS theory to access the strong coupling regime. We compute again, using analytical methods, the expectation value of $R^2$ and show how it interpolates smoothly between weak and strong coupling, and signals the phase transition in a manner consistent with previous numerical methods~\footnote{This phase transition is rather challenging to quantify. The order of the transition for low dimension $D$ is believed to be first, but becomes second order at higher dimensions. See\cite{Mandal:2009vz}-\cite{Morita:2020liy} for details. Our analysis is not precise enough to distinguish between the two cases.}. Our computation is however much more concise and is analytic up to the last step where one needs to find the roots of a polynomial. The technique is a demonstration of the power and promise of stochastic quantization as applied to BFSS Matrix theory, as well as to other settings involving holographic duality. In the Conclusion and Outlook section at the end, we comment how this technique can be generalized to cases involving fermions and supersymmetry, as well as other holographic settings -- specially when they exhibit chaotic dynamics at strong coupling. We also review alternate numerical techniques based on stochastic quantization that can be used for computing zero temperature dynamics.

In Section~\ref{sec:quant}, we develop the theory of stochastic quantization of the bosonic BFSS Matrix theory. In Section~\ref{sec:pert}, we compute the vev of $R^2$ perturbatively using stochastic Feynman diagrams. In Section~\ref{sec:var}, we develop the variational approach and compute the same quantity at all effective couplings, identifying the confinement/de-confinement phase transition. In Section~\ref{eq:num}, we present a numerical technique inspired from stochastic quantization. Finally, in Section~\ref{sec:conclusion}, we summarize the results and suggest directions for the future.

\section{Stochastic quantization of Matrix theory}
\label{sec:quant}

Consider the bosonic BFSS Matrix theory~\cite{Banks:1996vh,Ydri:2017ncg,Taylor:2001vb} with an added mass term, described by the Lagrangian
\begin{equation}\label{eq:action}
	S_E = \int d\tau\, \Tr \left(
	\frac{1}{2} \left(D_\tau X^i\right)^2-\frac{g^2}{4} \left[X^i,X^j\right]^2+\frac{\mu^2}{2} \left(X^i\right)^2
	\right)\ .
\end{equation}
The $X^i$'s are $N\times N$ matrices in the adjoint of $SU(N)$ with $i=1,\ldots D$, where $D\leq 9$. The mass term $\mu$ is needed to lift the flat directions that arise from commuting matrix configurations. This allows for stable lumps of energy centered at the origin. The coupling is denoted by $g$, which has units of energy cubed. At large $N$, the effective coupling is $\lambda\equiv g^2 N/E^3$ where $E$ is the energy scale of interest. We are working with the Euclidean form of the action using Euclidean time $\tau$ and the covariant derivative defined as
\begin{equation}
	D_\tau X^i = \partial_\tau X^i - i\,g\,\left[A, X^i\right]
\end{equation}
with $A(\tau)$ being the gauge field\footnote{From the standard Minkowski action, we change to $t\rightarrow -i\,t$, $A\rightarrow i\,A$ with the Minkowski and Euclidean actions related by $S=i\,S_E$ as usual.}. We want to study the quantum mechanics of this system through stochastic quantization~\cite{Damgaard:1988nq}, a technique that will give us access to the strong coupling regime.

In stochastic quantization, we think of the fields as being dependent on an additional `time' variable we call Langevin time $t$
\begin{equation}
	A(\tau)\rightarrow A(\tau,t)\ \ \ ,\ \ \ 
	X^i(\tau)\rightarrow X^i(\tau,t)\ ,
\end{equation}
and arrange for a stochastic evolution in this time direction given by the Langevin equations
\begin{equation}
	\frac{dX^i_{ab}}{dt} = - \frac{\delta S_E}{\delta X^i_{ba}} + \eta^i_{ab}(\tau,t)
\end{equation}
\begin{equation}
	\frac{dA_{ab}}{dt} = - \frac{\delta S_E}{\delta A_{ba}} + \eta_{ab}(\tau,t)\ .
\end{equation}
Here, $\eta^i(t)$ and $\eta(t)$ are Gaussian random noise variables in the adjoint of $SU(N)$ with statistical moments
\begin{equation}\label{eq:moments1}
	\left<\eta^i_{ab}\right> = 0\ \ \ ,\ \ \ \left<\eta_{ab}\right> = 0
\end{equation}
\begin{equation}
	\left<\eta^i_{ab}(\tau,t) \eta^j_{cd}(\tau',t')\right> = 2\, \delta_{ad} \delta_{bc} \delta^{ij} \delta(\tau-\tau') \delta(t-t')
\end{equation}
\begin{equation}
	\left<\eta_{ab}(\tau,t) \eta_{cd}(\tau',t')\right> = 2\, \delta_{ad} \delta_{bc} \delta(\tau-\tau') \delta(t-t')
\end{equation}
\begin{equation}\label{eq:moments2}
	\left<\eta^i_{ab}(\tau,t) \eta_{cd}(\tau',t')\right> = 0\ .
\end{equation}
The system then evolves in such a way so as to reach equilibrium at $t\rightarrow\infty$ when $\delta S_E/\delta X^i=\delta S_E/\delta A=0$. At this equilibrium, the fields fluctuate about the classical configuration, exploring randomly the phase space in a manner that mimics quantum fluctuations. This allows one to capture -- through stochastic fluctuations at equilibrium -- the phase space that dominates the quantum path integral of the system. As a result, we can compute the vev of any operator $\mathcal{O}(\tau)$ using
\begin{equation}
	\langle \mathcal{O}(\tau)\rangle = \lim_{t\rightarrow\infty} \langle \mathcal{O}(\tau, t)\rangle_\eta \equiv \lim_{t\rightarrow\infty} \frac{\int \mathcal{D}\eta\,\,\mathcal{O}(\tau, t) \exp\left[{-\frac{1}{4}\int d\tau dt\, \mbox{Tr}\,\eta^2(\tau,t)}\right]}{\int \mathcal{D}\eta\,\exp\left[{-\frac{1}{4}\int d\tau dt\, \mbox{Tr}\,\eta^2(\tau,t)}\right]}\ .
\end{equation}

One of the advantages of the stochastic quantization scheme when applied to a gauge theory arises from the fact that gauge fixing is not necessary when computing  expectation values of gauge invariant operators~\cite{Parisi:1980ys,Namiki:1982mj}. Yet, it is often still computationally helpful to implement `stochastic gauge fixing' that adds a frictional term to the stochastic evolution equations to drive the evolution dynamically towards a gauge slice of interest. This is done through a $t$-dependent gauge transformation. In general, the gauge transformations of our BFSS system are given by
\begin{equation} 
    \label{eq:gauge_transformations}
	A\rightarrow U^\dagger A U + \frac{i}{g} U^\dagger \partial_\tau U\ \ \ ,\ \ \ 
	X^i\rightarrow U^\dagger X^i U\ .
\end{equation}
We then choose a transformation $U(t)$ that depends on the Langevin time $t$, and write
a new transformed set of fields as~\cite{Namiki:1982mj}
\begin{equation}
	B = U^\dagger A U + \frac{i}{g} U^\dagger \partial_\tau U\ \ \ ,\ \ \ Y^i =  U^\dagger X^i U\ .
\end{equation}
We then define $\Lambda(t)$ through
\begin{equation}
	\frac{dU	}{dt} = g\,\frac{d\Lambda}{dt}U\ ,
\end{equation}
and choose
\begin{equation}
	\frac{d\Lambda}{dt} = -i\, \partial_\tau B\ ,
\end{equation}
which drives the stochastic evolution towards the Lorentz gauge condition
\begin{equation}
	\partial_\tau B=0\ .
\end{equation}
We then write the modified stochastic equations
\begin{equation}\label{eq:gaugedeq1}
	\frac{dY^i_{ab}}{dt} = - \frac{\delta S_E}{\delta Y^i_{ba}} + \eta^i_{ab}(\tau,t) + g\left[Y^i,\frac{d\Lambda}{dt}\right]_{ab}\ ,
\end{equation}
\begin{equation}\label{eq:gaugedeq2}
	\frac{dB_{ab}}{dt} = - \frac{\delta S_E}{\delta B_{ba}} + \eta_{ab}(\tau,t) + i\, D_B \left(\frac{d\Lambda}{dt}\right)_{ab}\ ,
\end{equation}
with
\begin{equation}
	g\left[Y^i,\frac{d\Lambda}{dt}\right] = -i\,g\, \left[Y^i, \partial_\tau B\right]\ ,
\end{equation}
\begin{equation}
	i\, D_B \left(\frac{d\Lambda}{dt}\right) = \partial_\tau^2 B - i\, g\, \left[B, \partial_\tau B\right]\ ,
\end{equation}
where we also note that
\begin{equation} \label{eq:noise_transformations}
	\eta \rightarrow U^\dagger \eta U\ ,
\end{equation}
which does not change the moments~(\ref{eq:moments1})-(\ref{eq:moments2}). The additional terms in~(\ref{eq:gaugedeq1}) and (\ref{eq:gaugedeq2}) effectively add corrections that typically would arise from the Faddeev-Popov method in traditional quantization, but no ghost fields are needed. 
As a result of all this, our system is described by the new Langevin equations
\begin{equation}\label{eq:eom1}
	\frac{dY^i}{dt} =  \partial_\tau^2 Y^i-\mu^2 Y^i+g^2 \left[Y^j,\left[Y^i, Y^j\right]\right]+2\,i\,g\,\left[\partial_\tau Y^i,B\right] - 2\,g^2\, \left[B,\left[Y^i,B\right]\right] + \eta^i(\tau,t)
\end{equation}
\begin{equation}\label{eq:eom2}
	\frac{dB}{dt} =  \partial_\tau^2 B -i\,g\, \left[\partial_\tau Y^i, Y^i\right] - 2\,g^2\, \left[Y^i, \left[B, Y^i\right]\right] - i\, g\, \left[B,\partial_\tau B\right]+ \eta(\tau,t)
\end{equation}
where we have used the action~(\ref{eq:action}). The new gauge fixing terms relax the stochastic evolution of the fields towards the $\partial_\tau B=0$ gauge slice. Solving these differential equations and taking $t\rightarrow\infty$ allows one to compute the quantum expectation value of any operator using statistical averages of the stochastically fluctuating fields at equilibrium. We will focus on computing a particular gauge invariant observable
\begin{equation}
	R^2 \equiv \frac{1}{N}\langle\mbox{Tr } {Y^i}^2\rangle
\end{equation}
where $R$ can be thought of as the physical size of the configuration in the $D$ dimensional target space. We will do this computation in three different ways: first, perturbatively in $g$; then, using a variational approach that works at strong coupling; and finally, using numerical techniques inspired from stochastic quantization that can also be extended to the strong coupling regime.

\section{Weak coupling}\label{sec:pert}

We start by introducing color indices 
\begin{equation}
	Y^i = Y^{i}_\alpha T^\alpha\ \ \ ,\ \ \ B = B_\alpha T^\alpha
\end{equation}
where the $T^\alpha$'s are the generators of $SU(N)$
\begin{equation}
	\mbox{Tr}\, T^\alpha T^\beta = \frac{1}{2} \delta^{\alpha\beta}\ \ \ ,\ \ \ \left[T^\alpha,T^\beta\right] = i f_{\alpha\beta\gamma}T^\gamma\ .
\end{equation}
We also Fourier transform in the direction of real time $\tau$ 
\begin{equation}
	B_\alpha(\tau, t) = \frac{1}{\sqrt{2\pi}}\int d\omega\, B_\alpha(\omega, t)\, e^{i\,\omega\,\tau}\ \ \ ,\ \ \ 	Y^i_\alpha(\tau, t) = \frac{1}{\sqrt{2\pi}}\int d\omega\, Y^i_\alpha(\omega, t)\, e^{i\,\omega\,\tau}\ .
\end{equation}
We then split the action into the free part $S^{(0)}_E$ and the interaction part $S^{(1)}_E$~\cite{Damgaard:1988nq}
\begin{equation}\label{eq:gaugedeq3}
	\frac{dY^i_{\alpha}(\Omega)}{dt} = - \frac{\delta S^{(0)}_E}{\delta Y^i_{\alpha}(\Omega)} - \frac{\delta S^{(1)}_E}{\delta Y^i_{\alpha}(\Omega)} + \eta^i_{\alpha}(\Omega,t)\ ,
\end{equation}
\begin{equation}\label{eq:gaugedeq4}
	\frac{dB_{\alpha}(\Omega)}{dt} = - \frac{\delta S{(0)}_E}{\delta B_{\alpha}(\Omega)} - \frac{\delta S^{(1)}_E}{\delta B_{\alpha}(\Omega)} + \eta_{\alpha}(\Omega,t)\ .
\end{equation}
where the variation of the free part is
\begin{equation}
	\frac{\delta S^{(0)}_E}{\delta Y^i_\alpha(\Omega)} = \left(\Omega^2+\mu^2\right)Y^i_\alpha(\Omega)\ ,
\end{equation}
\begin{equation}\label{eq:kinB}
	\frac{\delta S^{(0)}_E}{\delta B_\alpha(\Omega)} = \Omega^2 B_\alpha(\Omega)\ .
\end{equation}
The interaction part yields
\begin{eqnarray}
	\frac{\delta S^{(1)}_E}{\delta Y^i_\alpha(\Omega)} &=  \frac{2\,i\,g}{\sqrt{2\pi}} &\int  d\omega_1 d\omega_2\,\Omega f_{\alpha\beta\gamma} B_\gamma(\omega_1)Y_\beta^i(\omega_2) \delta(\omega_1+\omega_2-\Omega) \nonumber \\
	&+ \frac{g^2}{2\pi}&\int  d\omega_1 d\omega_2d\omega_3\,f_{\alpha\beta\gamma}f_{\delta\epsilon\gamma}Y_\delta^i(\omega_1)  \nonumber \\
	& &\times \left(Y^j_\epsilon(\omega_2)Y^j_\beta(\omega_3)-B_\epsilon(\omega_2)B_\beta(\omega_3)\right)\delta(\omega_1+\omega_2+\omega_3-\Omega)\ ,\label{eq:int1}
\end{eqnarray}
\begin{eqnarray}
	\frac{\delta S^{(1)}_E}{\delta B_\alpha(\Omega)} &=& \frac{i\, g}{\sqrt{2\,\pi}}\, \int d\omega_1 d\omega_2\,\omega_2\, f_{\alpha\beta\gamma}\left(Y^i_{\beta}(\omega_1) Y^i_{\gamma}(\omega_2)+i B_\beta(\omega_1) B_\gamma(\omega_2) \right)  \delta(\omega_1+\omega_2-\Omega) \nonumber \\
	&+& \frac{g^2}{2\pi}\int d\omega_1 d\omega_2d\omega_3\, f_{\alpha\beta\gamma}f_{\delta\epsilon\gamma}B_\delta(\omega_1) Y^i_\beta(\omega_2)Y^i_\epsilon(\omega_3)\delta(\omega_1+\omega_2+\omega_3-\Omega)\label{eq:int2}
\end{eqnarray}
using~(\ref{eq:action}), (\ref{eq:eom1}), and~(\ref{eq:eom2}). Note that these variations of $S^{(0)}_E$ and $S^{(1)}_E$ include the gauge fixing dissipative terms from (\ref{eq:gaugedeq1}) and~(\ref{eq:gaugedeq2});
in particular, this is the origin of the kinetic term for $B$ arising in~(\ref{eq:kinB}). Furthermore, the dissipative terms appear only in the variation of the action and do not have a representation in the action.

The solution to the stochastic equations~(\ref{eq:gaugedeq3}) and~(\ref{eq:gaugedeq4}) is then given by
\begin{equation}\label{eq:sol1}
	Y^i_\alpha(\Omega,t) = \int_0^t dt' G(\Omega, t-t')\left(
		\eta_\alpha^i(\Omega,t') - \frac{\delta S^{(1)}_E}{\delta Y^i_\alpha(\Omega,t')}
	\right)\ ,
\end{equation}
\begin{equation}\label{eq:sol2}
	B_\alpha(\Omega,t) = \int_0^t dt' G_B(\Omega, t-t')\left(
		\eta_\alpha(\Omega,t') - \frac{\delta S^{(1)}_E}{\delta B_\alpha(\Omega,t')}
	\right)\ ,
\end{equation}
where the Green functions must satisfy
\begin{equation}
	\frac{dG(\Omega,t)}{dt} = -\left(\Omega^2+\mu^2\right)G(\Omega,t)+\delta(t)\ \ \ ,\ \ \ \frac{dG_B(\Omega,t)}{dt} = -\Omega^2\,G_B(\Omega,t)+\delta(t)
\end{equation}
with boundary condition $G=G_B=0$ for $t<0$, with the stochastic evolution starting with random initial conditions at $t=0$. These are easily solved and yield~\cite{Damgaard:1988nq}
\begin{equation}
	G(\Omega, t-t') = e^{-\left(\Omega^2+\mu^2\right)(t-t')} \theta(t-t')\ ,
\end{equation}
\begin{equation}
	G_B(\Omega, t-t') = e^{-\Omega^2(t-t')} \theta(t-t')\ .
\end{equation}
Equations~(\ref{eq:sol1}) and~(\ref{eq:sol2}) can now be solved iteratively as an expansion in $g$. To one-loop order, we substitute~(\ref{eq:int1}) and~(\ref{eq:int2}) in~(\ref{eq:sol1}) and~(\ref{eq:sol2}) expanding to order $g^2$; this can then be substituted in
\begin{equation}
	\left<Y^i_\alpha(\omega)Y^i_\alpha(\omega')\right>
\end{equation}
to compute $R^2$ using~(\ref{eq:moments1})-(\ref{eq:moments2}) at one-loop order. The process can be organized diagrammatically, using so-called stochastic Feynman diagrams. 
\begin{figure}
\begin{center}
	\includegraphics[width=5in]{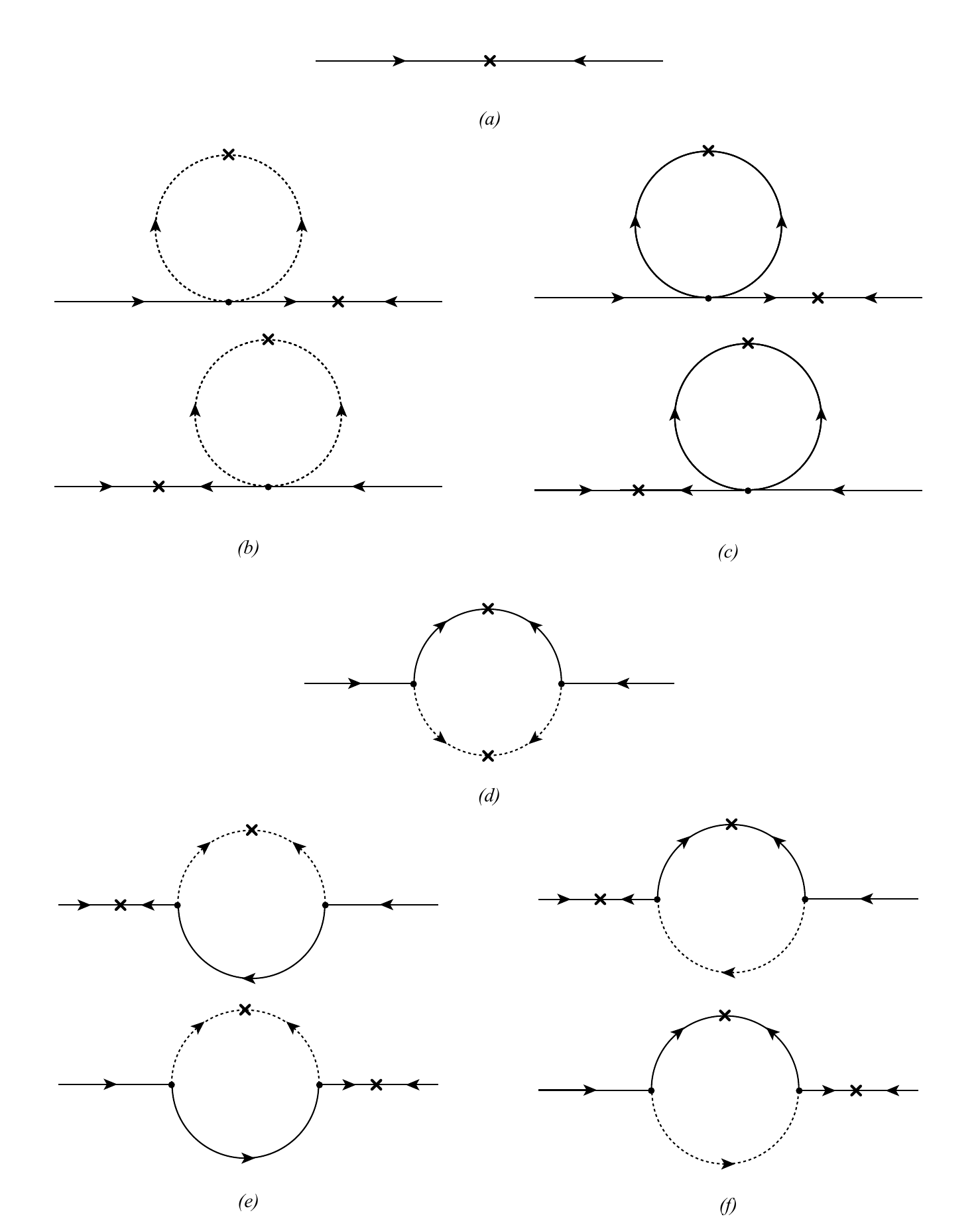}	
\end{center}
\caption{Stochastic Feynman diagrams for the computation of $R^2$ at one loop. Solid lines represent the propagator $G$ while dashed lines represent $G_B$. The `$\times$' represents insertion of a random noise. Three-pronged vertices come with a power of $g$, while four-pronged ones come with $g^2$. The Feynman rules are summarized in Appendix A.}\label{fig:Feynman}
\end{figure}
Figure~\ref{fig:Feynman} depicts all Feynman diagrams at play at one-loop order in computing $R^2$. We write the  result as
\begin{equation}
	\left<Y^i_\alpha(\omega)Y^i_\alpha(\omega')\right> = C^{(a)}\delta(\omega+\omega')+g^2 \sum_{I} C^{(I)}\delta(\omega+\omega')
\end{equation}
where
\begin{equation}
	C^{(a)} = \frac{D\,N^2}{\omega^2+\mu^2} \left(1-e^{-2\,t(\omega^2+\mu^2)}\right) \rightarrow \frac{D\,N^2}{\omega^2+\mu^2}
\end{equation}
comes from the zeroth order diagram in Figure~\ref{fig:Feynman}(a), after taking the $t\rightarrow\infty$ limit.
The $C^{(I)}$'s are organized in terms of five contributions, with $I\in \{b,c,d,e,f\}$. The first non-trivial diagrams are the tadpoles with the gauge field in the loop as shown in Figure~\ref{fig:Feynman}(b). One gets
\begin{equation}
	C^{(b)} = 0
\end{equation}
which arises through a cancellation in an expression of the form
\begin{equation}
	\lim_{t\rightarrow\infty}\int_{-\infty}^\infty d\omega\,d\omega_1\,\frac{1}{\omega_1^2}\left(\frac{1}{\omega^2+\mu^2}-\frac{e^{-2\,t\,\omega_1^2}}{\omega^2-\omega_1^2+\mu^2}\right) \rightarrow 0\ .
\end{equation}
Figure~\ref{fig:Feynman}(c) leads to the first non-zero contribution
\begin{equation}
	C^{(c)} = -\frac{4}{\pi} \int_{-\infty}^\infty d\omega_1\,\frac{D\,N\,(N^2-1)}{(\omega^2+\mu^2)^2(\omega_1^2+\mu^2)^2} = -\frac{4}{\mu} \frac{D\,N\,(N^2-1)}{(\omega^2+\mu^2)^2}
\end{equation}
where we used
\begin{equation}
	f_{\alpha\beta\gamma}f_{\alpha\beta\gamma} = N(N^2-1)\ .
\end{equation}
Another set of related diagrams shown in Figure~\ref{fig:Feynman}(d) give
\begin{equation}
	C^{(d)} = \frac{D\,N\,(N^2-1)\omega^2}{\mu\,(\omega^2+\mu^2)^4}\left( \frac{2\omega^2+3\mu^2}{\sqrt{3\omega^2+4\mu^2}}\mu+\omega^2-3\mu^2\right)
\end{equation}
while the diagrams in Figure~\ref{fig:Feynman}(e) give
\begin{equation}
	C^{(e)} = - \frac{D\,N\,(N^2-1) \omega^2(2\omega^2+3\mu^2)}{(\omega^2+\mu^2)^4\sqrt{3\omega^2+4\mu^2}}\ .
\end{equation}
Finally, from Figure~\ref{fig:Feynman}(f), the 
IR divergences are subtracted in the $t\rightarrow\infty$ limit as follows\footnote{This can be seen by adding an IR cutoff to the integrands and expanding in this cutoff. The subtlety has to do with the exchange of the order of two limits, that of taking the cutoff to zero and that of taking $t$ to infinity.}
\begin{eqnarray}
	&&\lim_{t\rightarrow\infty}\int_{-\infty}^\infty \frac{d\omega_1 d\omega}{\omega_1^2}\left(\frac{1}{(\omega^2+\mu^2)^2+(\omega^2+2\mu^2)\omega_1^2+\omega_1^4}-\frac{e^{-2\,t\,\omega_1^2}}{(\omega^2+\mu^2)^2-\omega^2\omega_1^2}\right) \nonumber \\
	&& =-\int_{-\infty}^\infty d\omega\, \frac{\pi (3\,\mu^2+2\,\omega^2)}{(\mu^2+\omega^2)^3\sqrt{3\,\omega^2+4\,\mu^2}}
\end{eqnarray}
and we get 
\begin{equation}
	C^{(f)} = D\,N\,(N^2-1)\frac{(\omega^2-\mu^2)}{2\,\mu\,(\omega^2+\mu^2)^3}+\frac{D\,N\,(N^2-1)\mu^2}{(\omega^2+\mu^2)^3\sqrt{3\,\omega^2+4\,\mu^2}}\ .
\end{equation}
Putting things together, we arrive at the one-loop result 
\begin{eqnarray}
	\left<Y^i_\alpha(\omega)Y^i_\alpha(\omega')\right> &&= \frac{D\,N^2}{\omega^2+\mu^2}\delta(\omega+\omega')\left[1 + \frac{g^2 N}{\omega^2+\mu^2}\right.\nonumber \\
	&&\left.\times\left(-\frac{7}{2\,\mu}-\frac{\mu}{\omega^2+\mu^2}+\frac{\omega^2(\omega^2-3\mu^2)}{\mu\,(\omega^2+\mu^2)^2}+\frac{\mu^2}{(\omega^2+\mu^2)\sqrt{3\omega^2+4\mu^2}}\right)\right]\ .
\end{eqnarray}
Fourier transforming back to $\tau$ space, we obtain
\begin{equation}
	R^2 = \frac{1}{N}\left<Y^i_\alpha (\tau)Y^i_\alpha(\tau)\right> = \frac{D\,N}{\mu}\left[\frac{1}{2}-\frac{g^2 N}{\mu^3} (N^2-1)\left(6\sqrt{3}+5\pi\right)\right]\ ,
\end{equation}
writing the one-loop correction in terms of the effective coupling $g^2 N/\mu^3$, measured at the scale of the mass term $\mu$ -- which functions as an IR cutoff or the inverse size of a box in which we put the configuration. We will next compute $R^2$ at any value of the effective coupling, including in the non-perturbative regime, using a powerful technique available in the stochastic quantization scheme known as the variational method~\cite{Grandati:1992hj}-\cite{Greensite:1982ck}.

\section{Variational technique and strong coupling}\label{sec:var}

The variational method~\cite{Greensite:1982ck,Damgaard:1988nq} relies on guessing an approximate form for the solution to~(\ref{eq:eom1}) and~(\ref{eq:eom2}), then minimizing a certain positive-definite quantity with respect to the parameters that describe the guess. If the guess is a good one, this method gives access to non-perturbative regimes. In the BFSS Matrix theory, we know that the chaotic nature of the dynamics at strong coupling arising from the quartic interaction in~(\ref{eq:action}) leads to an effective action of the form~\cite{Filev:2015hia}
\begin{equation}\label{eq:Seff}
	S_{eff} = \int d\tau\, \mbox{Tr}\left[\frac{1}{2}\left(\dot{Y}^i\right)^2+\frac{\sigma^2}{2} \left({Y}^i\right)^2\right]
\end{equation}
where $\sigma=\sigma(g)$ is an effective mass that is dependent on the coupling $g$. We then start with the ansatz
\begin{equation}
	\overline{Y}^i_\alpha(\Omega,t) = \int dt' G(\sigma|\Omega, t-t')\eta^i_\alpha(\Omega,t')\ \ \ ,\ \ \ \overline{B}^i_\alpha(\Omega,t) = \int dt' G_B(\Omega, t-t')\eta_\alpha(\Omega,t')
\end{equation}
where we define
\begin{equation}
	G(\sigma|\Omega, t-t') = e^{-\left(\Omega^2+\sigma^2\right)(t-t')} \theta(t-t')\ \ \ ,\ \ \ G_B(\Omega, t-t') = e^{-\Omega^2(t-t')} \theta(t-t')\ .
\end{equation}
This implies that
\begin{equation}\label{eq:varres}
	\frac{1}{N}\left<\overline{Y}^i_\alpha(\omega,t)\overline{Y}^j_\alpha(\omega',t)\right> = \frac{\delta^{ij} N}{\omega^2+\sigma^2}\delta(\omega+\omega')\Rightarrow
	R^2 = \frac{N\,D	}{2\,\sqrt{\sigma}}
\end{equation}
meaning that the system is effectively described by a free theory with a new effective mass $\sigma$. We then define the variational potential as~\cite{Greensite:1982ck}
\begin{equation}
	V(\sigma)\equiv \frac{1}{D\,N^2}\int_{-\infty}^\infty d\Omega \left<\left(\overline{Y}^i_\alpha(-\Omega,t)-{Y}^i_\alpha(-\Omega,t)\right)\left(\overline{Y}^i_\alpha(\Omega,t)-{Y}^i_\alpha(\Omega,t)\right)\right>\ ,
\end{equation}
where
\begin{equation}
	Y^i_\alpha(\Omega,t) = \int_0^t dt' G(\Omega, t-t')\left(
		\eta_\alpha^i(\Omega,t') - \frac{\delta S^{(1)}_E}{\delta \overline{Y}^i_\alpha(\Omega,t')}
	\right)\ .
\end{equation}
Minimizing $V(\sigma)$ with respect to $\sigma$ identifies the effective mass $\sigma=\sigma_0$ for any value of the coupling. Computing this potential analytically, we arrive at the expression
\begin{eqnarray}
	V(\sigma) &&= \left(1+\frac{1}{\sqrt{\sigma
   }}-\frac{2 \sqrt{2}}{\sqrt{\sigma +1}}\right)  \times \nonumber \\
   &&\left(\frac{1}{2}-\frac{g^2\left({N}^2-1\right)}{N}\frac{({D}-1)  }{(\sigma -1)
   \sqrt{\sigma }} +{g^4 \left({N}^2-1\right)}\frac{\left(4 ({D}-2) {D}^2+{D}+3\right) }{8 {D}(\sigma -1)^2 \sigma }\right) \nonumber \\
&&+{g^4 N^2}\frac{ \left({D}^2 \left(6 {N}^2-5\right)-5 {N}^2+6\right)}{\pi ^2 N^2 {D}}\,
I(\sigma)\ ,\label{eq:V}
\end{eqnarray}
with
\begin{eqnarray}
	I(\sigma) && \equiv \int_{-\infty}^\infty dp\,dp' \frac{1}{\left(p^2+1\right) \left({p'}^2+\sigma \right) \left(2 \sqrt{\sigma } \sqrt{2\,f_2+f_1+6 \sigma +2}+f_2+f_1+5 \sigma
   +1\right)}\times \nonumber \\
&& \left(\frac{1}{\left(f_1+4 \sigma \right) \sqrt{2
   f_2+f_1+6 \sigma +2}}+\frac{1}{4 \sqrt{\sigma }
   \left(f_1+4 \sigma \right)}\right)\ ,
\end{eqnarray}
\begin{equation}
	f_1\equiv \left(p+{p'}\right)^2\ \ \ ,\ \ \ f_2\equiv p^2+{p'}^2
\end{equation}
where we used~(\ref{eq:moments1})-(\ref{eq:moments2}) to compute the expectation value using stochastic averaging.
In this expression, we have applied a rescaling of $\tau$ and the fields so that $\mu=1$; equivalently, we measure the coupling in units of mass\footnote{Note that the rescaling $\tau'=\tau/a$, $B'=a\,B$, and $Y'=Y \sqrt{a}$ gives ${g'}^2=g^2 a^3$ and ${\mu'}^2=\mu^2 a^2$; choose $a=1/\mu$, we can set the mass parameter to one, measuring the effective coupling in units of $\mu$.} $\mu$ through the effective coupling $\lambda= g^2 N/\mu^3$. The expression given in~(\ref{eq:V}) is exact -- no perturbative expansion has been used in arriving at this expression. It is worthwhile noting that the computation involves the same IR divergences encountered in the perturbative treatment. However, with the same pattern encountered in the previous section, all divergence get cancelled in the limit of infinite Langevin time $t\rightarrow\infty$. The general scheme of these cancellations is as follows: for every IR divergent term which we write qualitatively as
\begin{equation}
	\frac{g(p)}{f(p)}
\end{equation}
where $f(p)\geq 0$ vanishes for zero energy $p=0$ but $g(p)$ remains finite, we find a term of the form
\begin{equation}
	-\frac{g(p) e^{-t f(p)}}{f(p)}\ .
\end{equation}
 In the limit $t\rightarrow\infty$, we write $e^{-t f(p)}\simeq 1$ near $f(p)=0$, cancelling the paired IR divergent term. More interestingly, the $g(p)$ in both expressions might be different but always matches near $f(p)=0$ as needed. Physically, the reason for this cancellation can be understood as follows: The IR divergence arises from the soft gluon running in loops; this is the non-physical gluon that has been made dynamical through stochastic gauge fixing. But in the $t\rightarrow\infty$ limit, the Langevin evolution has frictional terms that drive things to a gauge slice, freezing the unphysical gluon. So, one should expect that the IR divergences cancel only in the $t\rightarrow\infty$, which is what we find. As is typical in quantum field theory, the cancellation of these infinities is somewhat delicate and subtle. 

\begin{figure}
\begin{center}
	\includegraphics[width=5in]{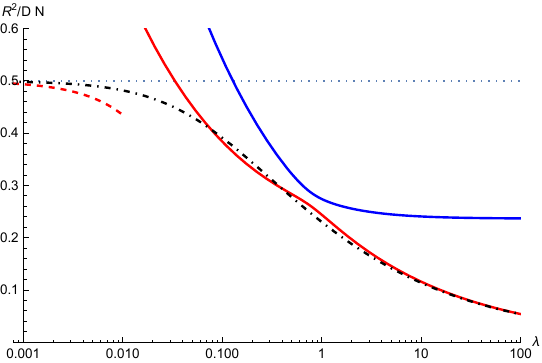}	
\end{center}
	\caption{$\xi\equiv R^2/D\,N$ as a function of effective coupling $\lambda$. The dashed line is the perturbative result from the previous section. The two real roots of~(\ref{eq:sixth}) are the predictions from the variational approach at asymptotically large $N$, depicted as solid lines (red and blue). The upper line (blue) is a false root. The dot-dashed black line is the result of finding the minimum of the full variational potential. We see the latter smoothly interpolates between weak and strong coupling. For this plot, we have $D=9$, $\mu=1$, and $N=30$.}\label{fig:min}
\end{figure}

The task is then to find the minimum of~(\ref{eq:V}) with respect to $\sigma$, solving $V'(\sigma)=0$. This is not a trivial problem, but there are several ways one can tackle the challenge. For example, in~\cite{Grandati:1993yc} a large $N$ expansion is developed that allows one to compute strong coupling corrections as an expansion in $1/N$. In our case, it is a rather straightforward task to analyze $V(\sigma)$ directly. Figure~\ref{fig:min} shows plots of $V(\sigma)$ for varied values of $N$, $\lambda$, and $D$. We see that a clear unique minimum can be identified in all cases. The general shape remains the same for all values of $N$. We note in particular that the minimum $\sigma_0$ becomes large with $N$. This means that large $N$ is also large $\sigma_0$. We have verified this monotonic relation by finding the minimum of the potential numerically: $\sigma_0 \propto N^\alpha$ for some positive $\alpha$.

The integral $I(\sigma)$ can be evaluated numerically. Furthermore, in the regime where $N$ is large, the numerical results show a clear robust asymptotic behavior
\begin{equation}
	I(\sigma) = \frac{1}{(e\,\sigma)^3}\ \ \ \mbox{Large $N$}
\end{equation}
where $e$ is the Euler number. Taking the large $N$ and $\sigma$ regime of~(\ref{eq:V}), we get
\begin{equation}
	V(\sigma) =  \frac{\lambda ^2}{\sigma^3}\frac{\left(6 {D}^2-5\right)}{e^3 \pi ^2 {D} }+\left(1+\frac{1}{\sqrt{\sigma }}-\frac{2
   \sqrt{2}}{\sqrt{\sigma }}\right) \left(\frac{1}{2}-\frac{\lambda }{\sigma
   ^{3/2}}{({D}-1) }+\frac{\lambda ^2}{\sigma ^3}\frac{\left(4 ({D}-2)
   {D}^2+{D}+3\right) }{8 {D} }\right)\ .
\end{equation}
To find the minimum $\sigma_0$, we need to solve $V'(\sigma)=0$; and we have from~(\ref{eq:varres})
\begin{equation}
	\sigma_0 = \frac{D^2 N^2}{4 R^4}\equiv \frac{1}{4\, \xi^2}\ ,
\end{equation}
where we defined $\xi\equiv R^2/D\,N$ for convenience. Finding $R^2$ then amounts to solving a sixth order polynomial in $\xi$ given by 
\begin{eqnarray}
	&&{28 \left(2 \sqrt{2}-1\right) \left(4 ({D}-2) {D}^2+{D}+3\right) }\lambda ^2
   \xi^6 \nonumber \\
   &&-\frac{12 \left(48 {D}^2+e^3 \pi ^2 \left(4 ({D}-2)
   {D}^2+{D}+3\right)-40\right)}{e^3 \pi ^2} \lambda ^2 \xi^5 \nonumber \\
   &&+8 \left(2-4
   \sqrt{2}\right) D ({D}-1)\, \lambda\, \xi^3+6 D ({D}-1)\, \lambda\, \xi^2+\frac{D}{\sqrt{2}}-\frac{D}{4}=0\ .\label{eq:sixth}
\end{eqnarray}
We have then distilled the problem of determining $R^2$ at strong coupling to an algebraic problem. Solving~(\ref{eq:sixth}) can be done numerically, and one finds {\em two} real roots for $\xi$. Figure~\ref{fig:min} shows the results. We collect several plots on the same graph: (1) The two real roots of~(\ref{eq:sixth}) are shown as solid curves; (2) We show the one-loop perturbative result from the previous section as a dashed line at weak coupling; (3) We also show the result of finding the minimum of~(\ref{eq:V}) numerically, without taking a large $N$ limit first -- and this is shown as a dot-dashed line that smoothly interpolates between weak coupling and one of the roots at strong coupling. We see that the variational approach  can capture decently the entire range of effective couplings $\lambda$, from small to large.  

Of the two large-$N$ asymptotic curves shown in the figure, the lower one (in red) that the dot-dashed curve hugs obviously describes the correct vacuum. To verify this further, we numerically compare the shape of the variational potentials before and after the large-$N$ asymptotic expansion, and one can verify that the upper curve in Figure~\ref{fig:min} corresponds to a maximum in the large-$N$ potential -- an extremum that does not exist in the general expression of the potential\footnote{The large $N$ asymptotic curve does not work well for small effective coupling $\lambda<0.1$, whereas the full variational potential does much better. We should not expect that the variational ansatz captures the physics at small coupling well. The surprising aspect here might be that the full form of the variational potential still works decently at weak coupling: this is because the variational ansatz is designed to also work exactly at zero effective coupling and, involving a smooth function, has to somehow interpolate in-between weak and strong. Taking the large $N$ limit drops the anchor of the ansatz at zero coupling, exposing more clearly that the variational ansatz's success at very small coupling is not to be taken very seriously.}. Yet another supporting argument for dropping the upper curve goes as follows: the vev of the action is proportional to $R^2$ -- using the virial theorem for the vacuum; hence, lower $R^2$ would be preferred, corresponding to the lower of the two asymptotic curves.  

In conclusion, our main result consist of the lower asymptotic curve (red) of Figure~\ref{fig:min} described implicitly by equation~(\ref{eq:sixth}), and the interpolating dot-dashed curve obtained from the minimum of the variational potential~(\ref{eq:V}) directly. These agree very well with the results from numerical Hamiltonian Mont\'{e} Carlo (HMC) computations in the literature~\cite{Filev:2015hia,Tanwar:2020fuv}\footnote{In the literature, the computations are done at finite temperature: one scans over the effective coupling by changing the temperature at fixed coupling $g$. In our case, we perform the computation at zero temperature but by varying $g$ while measuring the effective coupling at the mass scale. The two approaches are equivalent, resulting in scanning over the effective coupling $\lambda$ in a similar way and identifying the phase transition at $\lambda\sim 1$.}. The variational approach, with the ansatz we have employed, seems to capture the strong coupling physics well -- in addition to providing for a smooth interpolation from weak to strong coupling. There are variations of the ansatz one could have explored to improve the results, but we see that there is no need for this: the simplest ansatz works very well. This is a remarkable result: we have been able to get an unprecedented level of control over the strong coupling regime of bosonic Matrix theory.

\section{Numerical approaches}\label{eq:num}

Yet another approach to accessing the strongly coupled regime of a theory through stochastic quantization involves the direct numerical integration of the stochastic differential equations~(\ref{eq:eom1}) and~(\ref{eq:eom2})~\cite{HMCalgorithms}. This technique is closely related to the state-of-the-art in Lattice QCD computations, the so-called HMC method~\cite{Tanwar:2020fuv}\cite{HMCtheory}-\cite{Banuls:2019rao} -- albeit the connection is not rigorously established. 

The HMC technique has been developed thoroughly for the BFSS Matrix model at finite temperature in~\cite{Filev:2015hia,Tanwar:2020fuv}. In these examples, the gauge field is given a non-trivial holonomy along the $\tau$ circle whose period is interpreted as inverse temperature. Then the HMC method is applied for efficiently sampling the fields in the quantum path integral at finite temperature. In our case, we want to work at zero temperature with trivial holonomy in the gauge field, yet imposing periodic boundary conditions along $\tau$, for convenience, with a large enough period to implement the zero temperature limit. Furthermore, we will implement numerical techniques while using the stochastic gauge fixing approach that we introduced earlier -- an approach that has not been explored in the Matrix theory literature. With these two new ingredients, zero temperature computation and stochastic gauge fixing, we will present two different numerical methods -- alternatives to traditional HMC -- and compare them: (1) Employing direct integration of the BFSS stochastic differential equations using the SOSRI2 algorithm\footnote{We use the Julia programming language [julialang.org] for high performance and parallelization over 4 CPU cores. The SOSRI2 algorithm is implemented in the Differential Equations package [diffeq.sciml.ai/stable/]. It is a stability-optimized adaptive strong order 1.5 and weak order 2.0 Ito solver.}; (2) An implementation of the integration algorithm with an added Mont\'{e} Carlo step after each small time step of integration -- this is known as MALA (Metropolis-adjusted Langevin algorithm), a variant of HMC\cite{10.2307/3318418,mangoubi2019nonconvex} that is more rigorously related to the stochastic quantization problem.

\begin{figure}
\begin{center}
	\includegraphics[width=3in]{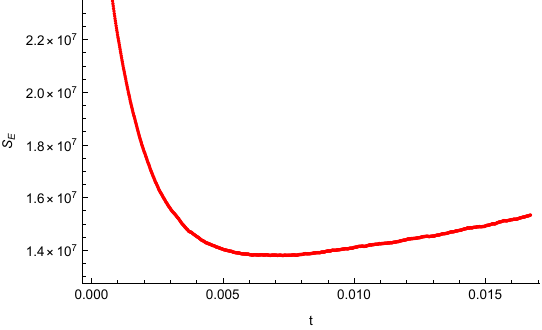}	
	\includegraphics[width=3in]{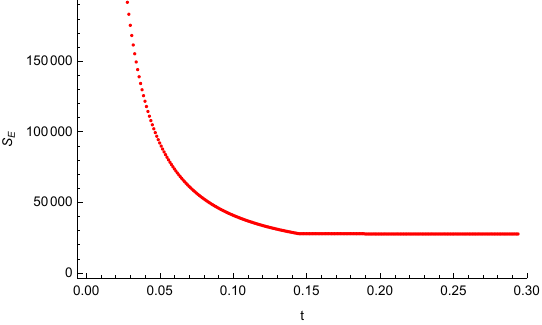}	
	
	(a)\hspace{3in}(b)
\end{center}
	\caption{(a) Failure of convergence of the Langevin evolution when using the SOSRI2 numerical integration, with $\lambda\simeq 1.0$. We are viewing the Euclidean action versus Langevin time; for larger Langevin times (not shown), the action diverges exponentially; (b) The MALA algorithm with $\lambda\simeq 1$ demonstrating quick and stable convergence. We have $N=10$ and $\mu=1$ for both graphs.}\label{fig:convergence}
\end{figure}

First, using direct integration through SOSRI2, we find that we are able to integrate the differential equations~(\ref{eq:eom1}) and~(\ref{eq:eom2}) for small enough effective coupling. However, Figure~\ref{fig:convergence}(a) shows the divergence of the method at strong effective coupling by plotting the Euclidean action as a function of Langevin time. After an initial exponential convergence from random initial conditions, the action starts to grow to infinity\footnote{In~\cite{nishimura2012origin}, it has been suggested that that a phenomenon similar to this is to be attributed to the physical emergence of three large space dimensions out of the total of nine. We do not necessarily see room for this interpretation in our case as we don't have enough statistics to draw this conclusion.}.
Hence, direct numerical integration of the stochastic differential equations does not work well. It would be interesting to look into the reasons for this as the direct integration method can be very useful as a check on other techniques.

\begin{figure}
\begin{center}
	\includegraphics[width=4in]{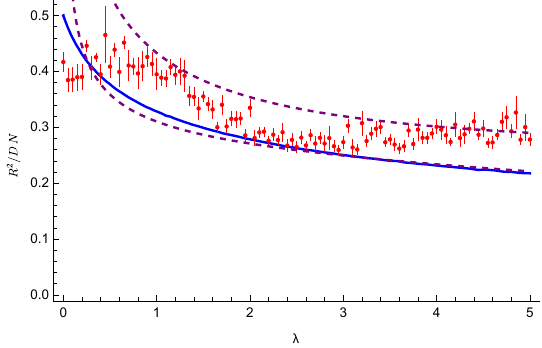}	
\end{center}
	\caption{$R^2/D\,N$ as a function of effective coupling $\lambda$, along with error bars. The dots show the simulation data using MALA. The curves are predictions from the variational approach. We have  $\mu=1$, $N=10$, $D=3$. The dashed lines are the two asymptotic variational results for large $N$. The data and curves for $\lambda\lesssim 1$ should not be relied on quantitatively.}\label{fig:mala}
\end{figure}

We then implement the MALA variant.  Figure~\ref{fig:convergence}(b) shows how this method leads to robust convergence. Indeed, the algorithm is extremely fast, with each evolution with $N=10$ taking a fraction of a second on a regular laptop. Using MALA, we computed the effective mass $\sigma_0$ as a function of effective coupling: We generated 10 or so evolutions per value of coupling to get an ensemble of field configurations with decent statistics; then we computed the correlation function
\begin{equation}
	\langle\mbox{Tr}\, Y^i(\tau)Y^(0)\rangle \propto e^{-\sigma_0 \tau}\ \ \ \mbox{For $\tau$ small enough}
\end{equation}
which allowed us to extract $\sigma_0$ using a fit at small times $\tau$. The reason for using small $\tau$ only is that our implementation has the $\tau$ living on a circle with a large enough period so as to reproduce zero temperature physics. We computed the correlation function using the standard Fast Fourier Transform method.
 Figure~\ref{fig:mala} shows the result from all this. We use~(\ref{eq:varres}) to compute $R^2$ since the computation of an effective mass assumes an effective action of the form given by~(\ref{eq:Seff}). Note that this is true only at large effective coupling; hence, the numerical data for small $\lambda$ should not be relied upon quantitatively. We see the transition between small and large effective coupling, which agrees with results from the literature using HMC. We also overlay on this data the results from the variational approach of the previous section.
The solid curve is the result of finding the minimum of the variational potential numerically. The two dashed curves are the two real roots of the polynomial~(\ref{eq:sixth}), valid if $N$ and $\sigma$ are very large.
We note again that it is not surprising that, at small effective coupling, the variational approach does not work well since the effective mass model is not a good description of the dynamics at weak coupling. So, in this figure, both MALA and the variational curves should be viewed only qualitatively at weak coupling. The variational approach however is reliable and very good at strong coupling (as compared to HMC techniques from the literature). We note however that MALA seems to oscillate between the two roots, the true vacuum at strong coupling and the false one. This is an interesting observation, perhaps suggesting the existence of a long-lived metastable vacuum at strong coupling at large $N$ -- at least long-lived enough for the stochastic evolution to get trapped in. Comparing with~\cite{Tanwar:2020fuv} where a traditional HMC numerical approach is used, we find that the HMC algorithm does better than MALA in that it tracks the variational asymptotic curve. MALA, as a very fast algorithm, can still be good for quickly exploring the parameter space, perhaps helping guide the construction of a variational ansatz. Quantitively however, it does not seem to work well-enough. The state-of-the-art HMC is still the most reliable numerical approach.

\section{Conclusion and outlook}\label{sec:conclusion}

In this work, we developed the stochastic quantization of the bosonic BFSS Matrix theory. The primary motivation was to access strong coupling computations more efficiently, and we have indeed demonstrated this. We captured the phase transition in the bosonic BFSS system that arises at large effective coupling. This transition, along with the behavior of the system in the strongly coupled phase, were determined using a variational technique, and the results agree with the numerical results in the literature that use HMC simulations. The variational technique however is more concise and stays analytical until the last step that involves finding the roots of a high order polynomial. It allows for much easier exploration of the parameter space of the theory, in addition to extracting asymptotic behaviors.

To develop these ideas further, one needs to extend the variational technique with two additional ingredients: the addition of fermions, and the use of the background field method to the compute effective actions. The first is needed to study the supersymmetric BFSS theory; and we know that supersymmetry plays a critical role in mapping the Matrix theory to eleven dimensional light-cone gravitational dynamics. The second is useful for computing various other observables that can help us understand the emergence of spacetime geometry from matrices. 

Fortunately, the inclusion of fermions in stochastic quantization has already been explored in the literature~\cite{Damgaard:1983tq}. Extending this to the BFSS Matrix theory would be straightforward. The best approach is not to integrate the fermions, as is often done in the literature, but to instead evolve the fermionic variables in Langevin time as Grassmanian numbers. The background field method has also been explored in the context of stochastic quantization~\cite{Okano:1986vr}. Adapting this to the BFSS Matrix model should also be possible. Variational methods in the context of stochastic quantization hold the potential for significant progress in our understanding of the BFSS conjecture and of emergent spacetime geometry. We hope to report on these directions shortly~\cite{WIP}. Other holographic dualities can benefit from this technology too. For example, in the AdS/CFT scenario, knowing the vev of operators from the gravitational computation gives us ways to develop robust ansatz for direct strong coupling computations using the variational technique. This can then be used to check the correspondence and compute systematically high curvature corrections to gravity. More generally, gravitational theories that often exhibit fast scrambling and chaotic dynamics offer certain simplifications for developing variational ansatz in the computation of vevs of operators in the dual strongly coupled theory.

On the numerical front, we presented two directions: (1) We showed that the direct integration of the stochastic equations does not work very well at strong coupling. If the direct integration algorithm can be refined without resorting to Mont\'{e} Carlo sampling, this would provide a new numerical technique for accessing strongly coupled dynamics that comes at the problem from a somewhat different angle than traditional techniques like HMC. This is challenging work that will remain for the future. (2) We demonstrated the use of the MALA algorithm in bosonic BFSS Matrix theory: it is computationally faster than HMC, but we also saw that it can be more prone to sampling errors. The MALA approach can be used to quickly explore the parameter space. Our viewpoint is that there is room for pursuing new numerical approaches beyond HMC -- within the context of stochastic quantization -- to complement the variational method that we showcased.

\section*{Acknowledgments}

This work was supported by the NSF grant numbered PHY-0968726 and the John Templeton Foundation grant 61149. VS would like to thank Saclay's IPhT for hosting him during the completion of this work.

\appendix

\newpage
\section{Appendix: Stochastic Feynman rules for BFSS Matrix theory}

\begin{figure}[h]
	\begin{center}
		\includegraphics[width=2.7in]{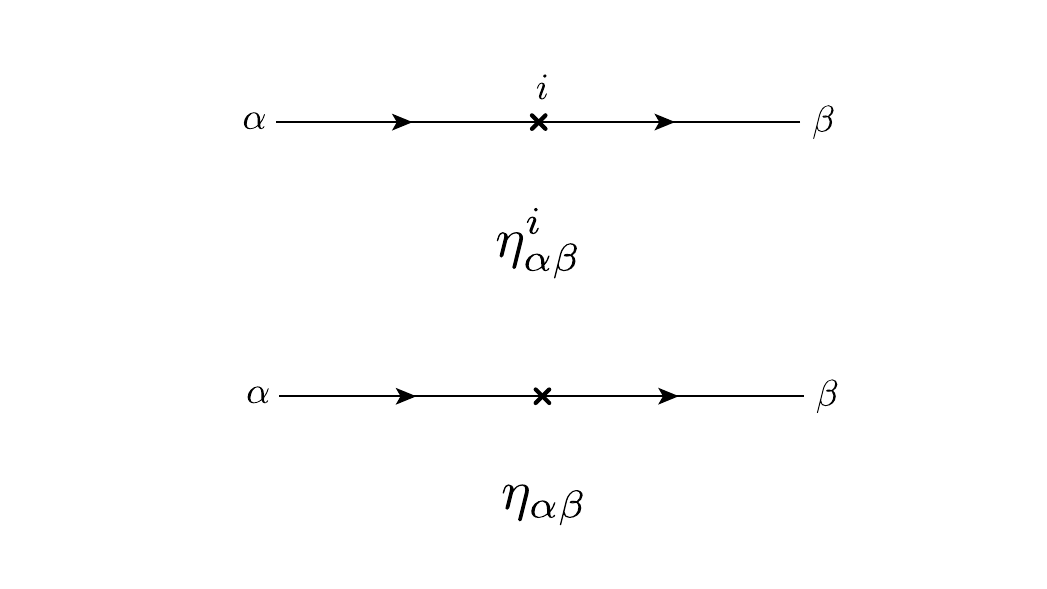}
		\includegraphics[width=2.7in]{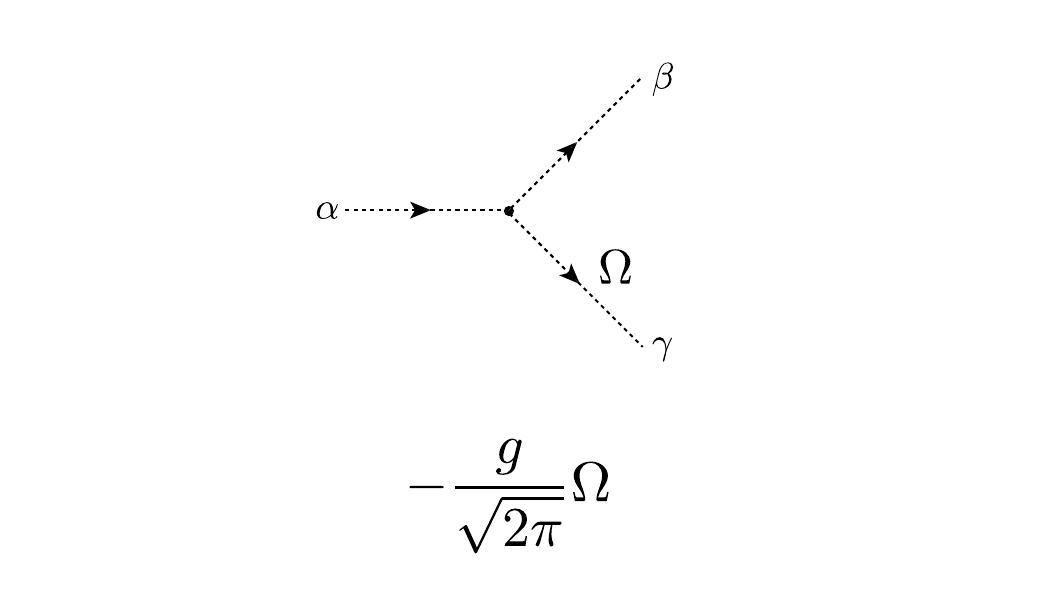}
	\end{center}
	\begin{center}
		\includegraphics[width=2.7in]{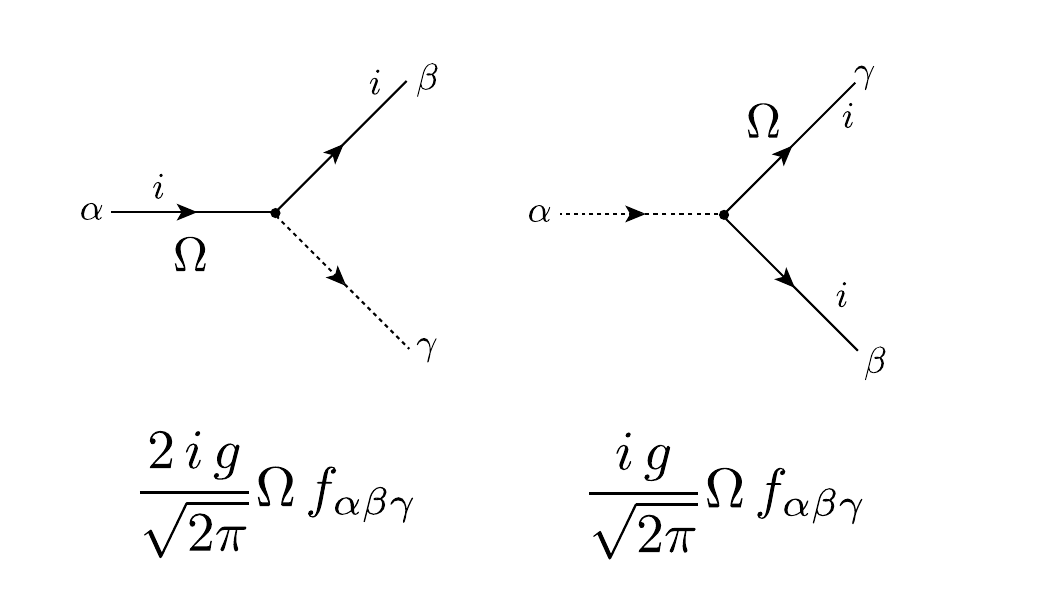}
		\includegraphics[width=2.7in]{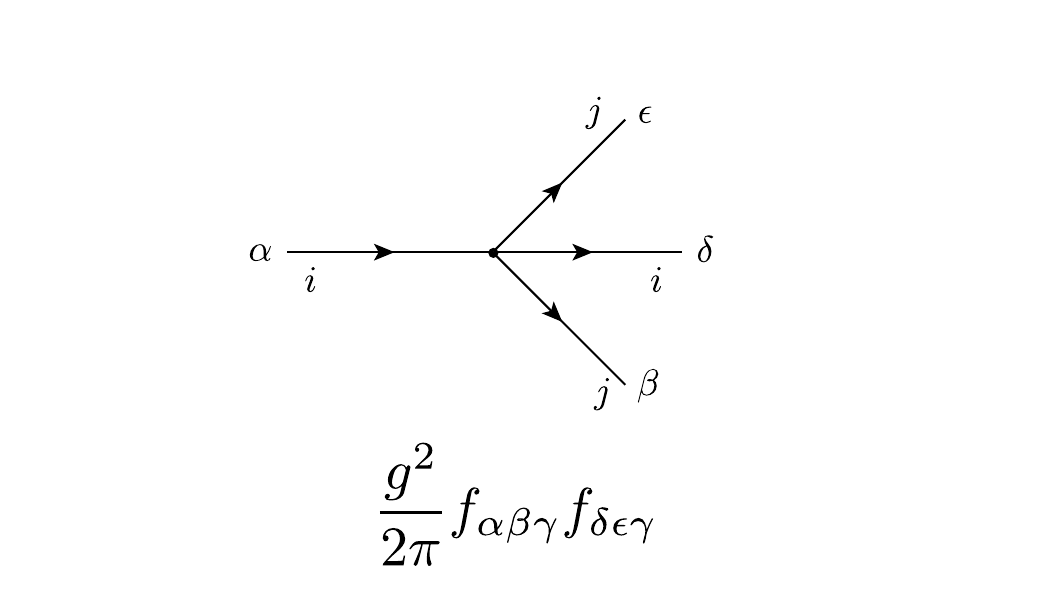}
	\end{center}
	\begin{center}
		\includegraphics[width=2.7in]{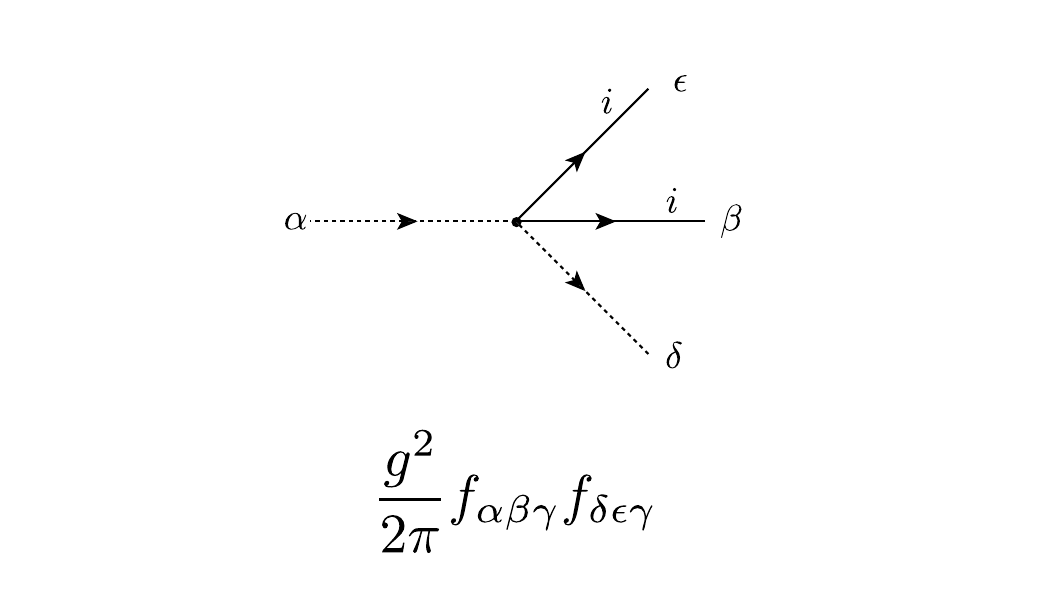}
		\includegraphics[width=2.7in]{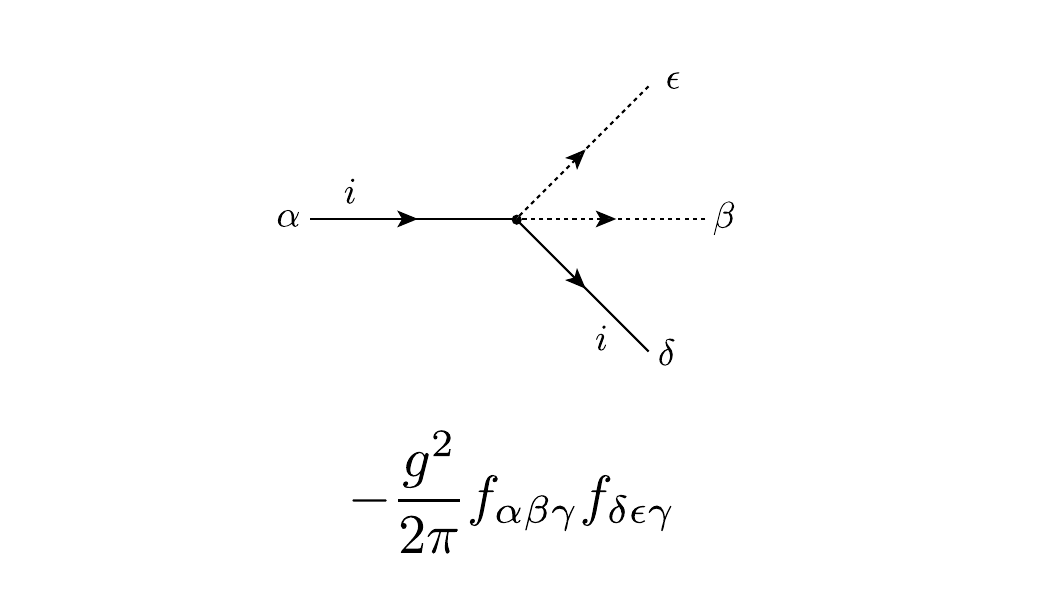}
	\end{center}
\end{figure}
These stochastic Feynman diagrams can be obtained from~(\ref{eq:int1}) and~(\ref{eq:int2}). Solid lines represent $Y$ fields ($G$ propagator), dashed lines represent the gauge field $B$ ($G_B$ propagator). Color indices are Greek letters as in the text. A cross represents an insertion of a noise $\eta$, which are then Wick-contracted while computing averages. $\Omega$ represents the momentum on a corresponding line. Note that stochastic diagrams have a distinguishing input line -- the line that comes in from the left. The rules of building diagrams from these are similar to the usual Feynman diagrams except for the insertion of noise crosses onto any propagator.

\bibliographystyle{utphys}
% \bibliography{biblio}

\providecommand{\href}[2]{#2}\begingroup\raggedright\endgroup

\end{document}